\documentclass[12pt]{article}
\usepackage{amssymb}
\usepackage{graphics}
\usepackage{epsfig}

\DeclareFontFamily{U}{rsf}{}
\DeclareFontShape{U}{rsf}{m}{n}{
  <5> <6> rsfs5 <7> <8> <9> rsfs7 <10-> rsfs10}{}
\DeclareMathAlphabet\Scr{U}{rsf}{m}{n}

\catcode`\@=11
\def\@citex[#1]#2{%
\if@filesw \immediate \write \@auxout {\string \citation {#2}}\fi
\@tempcntb\m@ne \let\@h@ld\relax \def\@citea{}%
\@cite{%
  \@for \@citeb:=#2\do {%
    \@ifundefined {b@\@citeb}%
      {\@h@ld\@citea\@tempcntb\m@ne{\bf ?}%
      \@warning {Citation `\@citeb ' on page \thepage \space undefined}}%
      {\@tempcnta\@tempcntb \advance\@tempcnta\@ne%
      \@tempcntb\number\csname b@\@citeb \endcsname \relax%
      \ifnum\@tempcnta=\@tempcntb 
        \ifx\@h@ld\relax%
          \edef \@h@ld{\@citea\csname b@\@citeb\endcsname}%
        \else%
          \edef\@h@ld{\ifmmode{-}\else--\fi\csname b@\@citeb\endcsname}%
        \fi%
      \else
        \@h@ld\@citea\csname b@\@citeb \endcsname%
        \let\@h@ld\relax%
      \fi}%
    \def\@citea{,\penalty\@highpenalty\,}%
  }\@h@ld
}{#1}}

\def\@citeb#1#2{{[#1]\if@tempswa , #2\fi}}
\def\@citeu#1#2{{$^{#1}$\if@tempswa , #2\fi }}
\def\@citep#1#2{{#1\if@tempswa , #2\fi}}

\def\bcites{         
        \catcode`\@=11
        \let\@cite=\@citeb
        \catcode`\@=12
}

\def\upcites{         
        \catcode`\@=11
        \let\@cite=\@citeu
        \catcode`\@=12
}

\def\plaincites{      
        \catcode`\@=11
        \let\@cite=\@citep
        \catcode`\@=12
}

\newcount\hour
\newcount\minute
\newtoks\amorpm
\hour=\time\divide\hour by 60
\minute=\time{\multiply\hour by 60 \global\advance\minute by-\hour}
\edef\standardtime{{\ifnum\hour<12 \global\amorpm={am}%
        \else\global\amorpm={pm}\advance\hour by-12 \fi
        \ifnum\hour=0 \hour=12 \fi
        \number\hour:\ifnum\minute<10 0\fi\number\minute\the\amorpm}}
\edef\militarytime{\number\hour:\ifnum\minute<10 0\fi\number\minute}

\def\draftlabel#1{{\@bsphack\if@filesw {\let\thepage\relax
   \xdef\@gtempa{\write\@auxout{\string
      \newlabel{#1}{{\@currentlabel}{\thepage}}}}}\@gtempa
   \if@nobreak \ifvmode\nobreak\fi\fi\fi\@esphack}
        \gdef\@eqnlabel{#1}}
\def\@eqnlabel{}
\def\@vacuum{}
\def\marginnote#1{}
\def\draftmarginnote#1{\marginpar{\raggedright\scriptsize\tt#1}}
\def\draft{
        \pagestyle{plain}
        \overfullrule=2pt
        \oddsidemargin -.5truein
        \def\@oddhead{\sl \phantom{\today\quad\militarytime} \hfil
        \smash{\Large\sl DRAFT} \hfil \today\quad\militarytime}
        \let\@evenhead\@oddhead
        \let\label=\draftlabel
        \let\marginnote=\draftmarginnote
        \def\ps@empty{\let\@mkboth\@gobbletwo
        \def\@oddfoot{\hfil \smash{\Large\sl DRAFT} \hfil}
        \let\@evenfoot\@oddhead}
        \def\@eqnnum{(\theequation)\rlap{\kern\marginparsep\tt\@eqnlabel}%
        \global\let\@eqnlabel\@vacuum}  }

\def\section{\@startsection {section}{1}{\z@}{3.ex plus 1ex minus
 .2ex}{2.ex plus .2ex}{\large\bf}}
\def\subsection{\@startsection{subsection}{2}{\z@}{2.75ex plus 1ex minus
 .2ex}{1.5ex plus .2ex}{\bf}}

\def\appendix{{\newpage\section*{Appendix}}\let\appendix\section%
        {\setcounter{section}{0}
        \gdef\thesection{\Alph{section}}}\section}

\def\abstract{\if@twocolumn
\section*{Abstract}
\else 
\begin{center}
{\bf Abstract\vspace{-.5em}\vspace{0pt}}
\end{center}
\quotation
\fi}

\catcode`\@=12

\newcommand{\beq}{\begin{equation}}
\newcommand{\eeq}{\end{equation}}
\newcommand{\beqa}{\begin{eqnarray}}
\newcommand{\eeqa}{\end{eqnarray}}

\newcommand{\Z}{{\bf Z}}

\newcommand{\R}{{\bf R}}

\newcommand{\be}{\begin{eqnarray}}
\newcommand{\ee}{\end{eqnarray}}
\newcommand{\nn}{\nonumber}

\def\lae{\mathrel{\mathop{\smash{\lower .5 ex \hbox{$\stackrel<\sim$}}}}}
\def\lae{\mathrel{\mathop{\smash{\lower .5 ex \hbox{$\stackrel>\sim$}}}}}

\def\vev#1{\left\langle #1 \right\rangle}

\def\Tr{{\rm Tr}}

\def\l:{\mathopen{:}\,}
\def\r:{\,\mathclose{:}}

\catcode`\@=11
\def\theequation{\arabic{equation}}
\catcode`\@=12

\bcites

\catcode`\@=11
\def\theequation{\thesection.\arabic{equation}}
\@addtoreset{equation}{section}
\@addtoreset{footnote}{section}
\@addtoreset{footnote}{subsection}
\catcode`\@=12

\newcommand{\ft}[2]{{\textstyle\frac{#1}{#2}}}
\newcommand{\eqn}[1]{(\ref{#1})}
\def\Dslash{\,\,{\raise.15ex\hbox{/}\mkern-12mu D}}
\def\Dbarslash{\,\,{\raise.15ex\hbox{/}\mkern-12mu {\bar D}}}
\def\delslash{\,\,{\raise.15ex\hbox{/}\mkern-9mu \partial}}
\def\delbarslash{\,\,{\raise.15ex\hbox{/}\mkern-9mu {\bar\partial}}}
\def\pslash{\,\,{\raise.15ex\hbox{/}\mkern-9mu p}}
\def\calDslash{\,\,{\raise.15ex\hbox{/}\mkern-12mu {\cal D}}}

\newcommand{\vg}{\vec{g}}
\newcommand{\valpha}{\vec{\alpha}}

\begin{document}
\pagestyle{plain}
\setcounter{page}{1}
\newcounter{bean}
\baselineskip16pt

\begin{titlepage}

\begin{center}

\hfill\today\\

\vskip 3.0 cm {\Large On Monopoles and Domain Walls} \vskip 1 cm
{Amihay Hanany${}^1$ and David Tong${}^2$}\\
\vskip 1cm
{\sl ${}^1$ Center for Theoretical Physics, Massachusetts Institute of
Technology, \\
Cambridge, MA 02139, USA.\\}
{\tt hanany@mit.edu}\\
\vskip .5cm
{\sl ${}^2$ Department of Applied Mathematics and Theoretical Physics, \\
University of Cambridge, CB3 0WA, UK.} \\
{\tt d.tong@damtp.cam.ac.uk}

\end{center}

\vskip 1.5 cm
\begin{abstract}
The purpose of this paper is to describe a relationship between
maximally supersymmetric domain walls and magnetic monopoles. We
show that the moduli space of domain walls in non-abelian gauge
theories with $N$ flavors is isomorphic to a complex, middle
dimensional, submanifold of the moduli space of $U(N)$ magnetic
monopoles. This submanifold is defined by the fixed point set of a
circle action rotating the monopoles in the plane. To derive this
result we present a D-brane construction of domain walls, yielding
a description of their dynamics in terms of truncated Nahm
equations. The physical explanation for the relationship lies in
the fact that domain walls, in the guise of kinks on a vortex
string, correspond to magnetic monopoles confined by the Meissner
effect.

\end{abstract}
\end{titlepage}

\tableofcontents

\section{Introduction}

Domain walls in gauge theories with eight supercharges have rather
special properties. These walls were first studied by Abraham and
Townsend \cite{edpaul} who showed that in two-dimensions, where
domain walls are known as kinks, they exhibit dyonic behaviour
reminiscent of magnetic monopoles. Further similarities between
kinks and  magnetic monopoles, at both the classical and quantum
level, were uncovered in \cite{nick}. The physical explanation for
this relationship  was presented in \cite{stillme}, where new BPS
solutions were described corresponding to magnetic monopoles in a
phase with fully broken gauge symmetry. The Meissner effect
ensures that monopoles are confined: the magnetic flux is unable
to propagate through the vacuum and leaves the monopole in two
collimated tubes. From the perspective of the flux tube, the
monopole appears as a kink. The idea of describing confined
monopoles as kinks in $Z_N$ strings occurred previously in
\cite{bead}. The relationship between the confined magnetic
monopoles and the kink was further explored in
\cite{shif,meami,auzzi} and related systems were studied in
\cite{o1,o2,o3,o4,o5,o6}.

In this paper we use D-brane techniques to study the moduli space
of multiple domain walls. This allows us to develop a description
of the domain wall dynamics in terms of a linearized Nahm
equation, providing a direct relationship to the dynamics of
monopoles. Specifically, we show that the moduli space of domain
walls, which we denote as ${\cal W}_{\vg}$, is isomorphic to a
middle dimensional submanifold of the moduli space of magnetic
monopoles ${\cal M}_{\vg}$. This submanifold describes magnetic
monopoles lying along a line, and can be described as the fixed
point of an ${\bf S}^1$ action $\hat{k}$, rotating the monopoles
in a plane,
\be {\cal W}_{\vg}\cong\left.{\cal
M}_{\vg}\right|_{\hat{k}=0}\label{resu}\ee
The correspondence captures the topology and asymptotic metric of
the domain wall moduli space ${\cal W}_{\vg}$. It does not extend
to the full metric on ${\cal W}_{\vg}$. Nevertheless, as we shall
explain, it does correctly capture the most important feature of
domain walls: their ordering along the line.

The relationship \eqn{resu} plays companion to the result of
\cite{vib}, where the moduli space of vortices was shown to be a
middle dimensional submanifold of the moduli space of instantons.
Indeed, upon dimensional reduction,  the self-dual instanton
equations become the monopole equations, while the vortex
equations descend to the domain wall equations.

We start in the following section by describing the domain walls
in question, together with a review of their moduli spaces. We pay
particular attention to the crudest physical feature of domain
walls, namely the rules governing their spatial ordering along the
line. Section 3  contains a brief review of magnetic monopoles in
higher rank gauge groups, primarily in order to fix notation, allowing us
to elaborate on the relationship \eqn{resu}. We also describe the
Nahm construction of the monopole moduli space as it arises from
D-branes. The meat of the paper is in Section 4. We present a
D-brane embedding of domain wall solitons which gives a
description of their dynamics in terms of a linear Nahm equation.
This equation is somewhat trivial, with the content hidden in
various boundary conditions. We show how these boundary conditions
capture the prescribed ordering of domain walls.

\section{Domain Walls}

In this paper we will study a class of BPS domain wall solutions occurring in
maximally supersymmetric theories with multiple, isolated vacua. The Lagrangian
for these models includes a $U(k)$ gauge field $A_\mu$, a real adjoint
scalar $\sigma$ and $N$ fundamental scalars $q_i$, each with real mass $m_i$
\be {\cal L} ={\Tr} \left[\frac{1}{4e^2} F_{\mu\nu}F^{\mu\nu}
+\frac{1}{2e^2}|{\cal D}_\mu\sigma|^2 +\frac{e^2}{2}(q_i\!\otimes
\!q_i^\dagger-v^2)^2\right] +\sum_{i=1}^{N}\left[|{\cal D}_\mu
q_i|^2 +q_i^\dagger(\sigma-m_i)^2q_i \right]\nn\ee
where there is an implicit sum over the flavor index $i$ in the
adjoint valued term $q_i\otimes q_i^\dagger$. This Lagrangian can
be embedded in a theory with 8 supercharges in any spacetime
dimension $1\leq d\leq 5$ (e.g. ${\cal N}=2$ SQCD in $d=3+1$).
Such theories include further scalar fields which can be shown to
vanish on the domain wall solutions\footnote{If we promote the scalar field
$\sigma$ and the masses $m_i$ to complex variables, then the theories admit an
interesting array of domain wall junctions \cite{web} and dyonic walls \cite{14}.}.
The fermions do contribute zero modes but will not be important here.

When the Higgs expectation value $v^2$ is non-vanishing, and the
masses $m_i$ are distinct ($m_i\neq m_j$ for $i\neq j$), the
theory has a set of isolated vacua. Each vacuum is labelled by a
set $\Xi$ of $k$ distinct elements, chosen from a possible $N$,
\be \Xi = \left\{\xi(a): \xi(a)\neq \xi(b)\ {\rm for}\
a\neq b\right\} \label{set}\ee
Here $a=1,\ldots,k$ runs over the color index, while $\xi(a)\in
\{1,\ldots,N\}$. Up to a gauge transformation, the vacuum
associated to this set is given by,
\be \sigma={\rm diag}(m_{\xi(1)},\ldots,m_{\xi(k)}) \ \ \ \
\ \ ,\ \ \ \ \ \ \ q^a_{\ i}=v\,\delta^a_{\ i=\xi(a)}
\label{vacuum}\ee
For $N<k$ there are no supersymmetric vacua; for $N\geq
k$, the number of vacua is
\be N_{\rm vac}={ N\choose k}=\frac{N!}{k!(N-k)!} \label{nvac}\ee
Each of these vacua is isolated and exhibits a mass gap. There
are $k^2$ non-BPS massive gauge bosons and quarks with masses
$m^2_\gamma\sim e^2v^2+|m_{\xi(a)}-m_{\xi(b)}|^2$ and
$k(N-k)$ BPS massive quark fields with masses
$m_q\sim|m_{\xi(a)}-m_i|$ (with $i\notin \Xi$).

For vanishing masses $m_i=0$ the theory enjoys an $SU(N)$ flavor
symmetry, rotating the $q_i$. When distinct masses are turned on
this is broken explicitly to the Cartan-sub-algebra $U(1)^{N-1}$.
Meanwhile, the $U(k)$ gauge group is broken spontaneously in the
vacuum by the expectation values \eqn{vacuum}.

The existence of multiple, gapped, isolated vacua is sufficient to
guarantee the existence of co-dimension one domain walls
(otherwise known as kinks). These walls are BPS objects,
satisfying first order Bogomoln'yi equations which can be derived
in the usual manner by completing the square. We first choose a
flat connection $F_{\mu\nu}=0$, and allow the fields to depend
only on a single coordinate, say $x^3$. Then the Hamiltonian is
given by
\be {\cal H}&=& {\Tr} \left[\frac{1}{2e^2}|{\cal D}_3\sigma|^2 +
\frac{e^2}{2}(q_i\!\otimes \!q_i^\dagger-v^2)^2\right] +
\sum_{i=1}^{N}\left[|{\cal D}_3 q_i|^2
+q_i^\dagger(\sigma-m_i)^2q_i \right]\nn
\\  &=& \frac{1}{2e^2}\Tr\left[{\cal D}_3\sigma - e^2
(q_i\!\otimes\! q_i^\dagger -v^2)\right]^2 + \sum_{i=1}^{N} |{\cal
D}_3q_i-(\sigma-m_i)q_i|^2 \nn\\ && +\ \Tr \left[ {\cal
D}_3\sigma\,(q_i\!\otimes\!q_i^\dagger-v^2) \right]
+\sum_{i=1}^{N} \left[ q_i^\dagger(\sigma-m_i){\cal D}_3q_i+{\cal
D}_3q_i^\dagger(\sigma-m_i)q_i\right]
\\\nn &\geq& -\partial_3\left(v^2\,\Tr\,\sigma\right)
\ee
Our domain wall interpolates between a vacuum $\Xi_-$ at
$x^3=-\infty$, as determined by a set \eqn{set}, and a distinct
vacuum ${\Xi}_+$ at $x^3=+\infty$. The minus signs above have been
chosen under the assumption that $\Delta m>0$, where $\Delta
m=\sum_{i\in\,{\Xi}_-}m_i-\sum_{i\in\,\Xi_+}m_i$, so that a BPS
domain wall satisfies the Bogomoln'yi equations,
\be {\cal D}_3\sigma=e^2(q_i\!\otimes\!q_i^\dagger -v^2) \ \ \ \ \
, \ \ \ \ \ {\cal D}_3q_i=(\sigma-m_i)q_i \label{bog}\ee
and has tension given by $T=v^2\Delta m$. Analytic solutions to
these equations can be found in the $e^2\rightarrow \infty$ limit
\cite{me,isozumi,isonon}, which give smooth approximations to the
solution at large, but finite $e^2$ \cite{finite}.

\subsection{Classification of Domain Walls}

Domain walls in field theories are classified by the choice of
vacuum $\Xi_-$ and ${\Xi}_+$ at left and right infinity. However,
our theory contains an exponentially large number of vacua
\eqn{nvac} and one may hope that there is a coarser, less
unwieldy, classification which captures certain relevant
properties of a given domain wall. Such a classification was
offered in \cite{boojum}.

Firstly define the $N$-vector $\vec{m}=(m_1,\ldots, m_{N})$. The tension of the
BPS domain wall can then be written as
\be T_{\vg}=v^2\Delta m\equiv v^2\vec{m}\cdot \vec{g}
\label{tvg}\ee where the $N$-vector $\vec{g}\in\Lambda_R(su(N))$,
the root lattice of $su(N)$. Note that there do not exist domain
wall solutions for all $\vec{g}\in \Lambda_R(su(N))$; the only
admissible vectors are of the form $\vec{g}=(p_1,\ldots,p_{N})$
with $p_i=0$ or $\pm 1$. Note also that a choice of $\vec{g}$ does
not specify a unique choice of vacua $\Xi_-$ and ${\Xi}_+$ at left
and right infinity. Nor, in fact, does it specify a unique domain
wall moduli space ${\cal W}_{\vg}$. Nevertheless, domain walls in
sectors with the same $\vg$ share common traits.

The dimension of the moduli space of domain wall solutions was
computed in \cite{boojum} using an index theorem, following earlier
results in \cite{lee,isonon}.
To describe the dimension of the moduli space, it is useful to
decompose
$\vec{g}$ in terms of simple roots\footnote{The basis of simple
roots is fixed by the requirement that $\vec{m}\cdot
\vec{\alpha}_i >0$ for each $i$. A unique basis is defined in this
way if $\vec{m}$ lies in a positive Weyl chamber, which occurs
whenever the masses are distinct so that $SU(N)\rightarrow
U(1)^{N-1}$. If we choose the ordering $m_1>m_2>\ldots > m_{N}$ we
have simple roots $\vec{\alpha}_1=(1,-1,0,\ldots,0)$ and
$\vec{\alpha}_2=(0,1,-1,0,\ldots,0)$ through to
$\vec{\alpha}_{N-1}=(0,\ldots,1,-1)$.} $\vec{\alpha}_i$,
\be
\vec{g} =\sum_{i=1}^{N-1}n_i\vec{\alpha}_i\ \ \ \  \ \ , \ \ \ \ \ \ n_i\in\Z
\label{dwg}\ee
The index theorem of
\cite{boojum} reveals that domain wall solutions to \eqn{bog}
exist only if $n_i\geq 0$ for all $i$. If this holds, the number
of zero modes of a solution is given by
\be
{\rm dim}\left({\cal W}_{\vec{g}}\right) = 2\sum_{i=1}^{N-1}n_i
\label{dwdim}\ee
where ${\cal W}_{\vg}$ denotes the moduli space of any set of
domain walls with charge $\vg$. This result has a simple physical
interpretation. There exist $N-1$ types of ``elementary'' domain
walls corresponding to a $\vec{g}=\vec{\alpha}_i$, the simple
roots. Each of these has just two collective coordinates
corresponding to a position in the $x^3$ direction and a phase. As
first explained in \cite{edpaul}, the phase coordinate is a
Goldstone mode arising because the domain wall configuration
breaks the $U(1)^{N-1}$ flavor symmetry as we review below. In
general, a domain wall labelled by $\vec{g}$ can be thought to be
constructed from $\sum_in_i$ elementary domain walls, each with
its own position and phase collective coordinate. Let us now turn
to some examples.

\subsection{The Moduli Space of Domain Walls: Some Examples}

\subsubsection*{\it Example 1: $\vg=\valpha_1$}

The simplest system admitting a domain wall is the abelian $k=1$ theory with
$N=2$ charged scalars $q_1$ and $q_2$. The $N_{vac}=2$ vacua of the
theory are given by $\sigma=m_i$ and $|q_j|^2=v^2\delta_{ij}$ for $i=1,2$.
There is a single domain wall in this theory with $\vec{g}=\vec{\alpha}_1$ interpolating
between these two vacua. Under the $U(1)_F$ flavor symmetry, $q_1$ has charge $+1$ and
$q_2$ has charge $-1$. In each of the vacua, the $U(1)_F$ symmetry coincides
with the $U(1)$ gauge action but, in the core of the domain wall, both $q_1$ and
$q_2$ are non-vanishing, and $U(1)_F$ acts non-trivially. The resulting goldstone
mode is the phase collective coordinate.
The moduli space of the domain wall is simply
\be {\cal W}_{\vec{g}=\vec{\alpha}_1}\cong\R\times{\bf S}^1
\ee
where the $\R$ factor describes the center of mass of the domain wall and
the ${\bf S}^1$ the phase. One can show that the ${\bf S}^1$ has radius
$2\pi v^2/T_{\vg}=2\pi /(m_1-m_2)$.

\subsubsection*{\it Example 2: $\vg=\valpha_1+\valpha_2$}

The simplest system admitting multiple domain walls is the abelian
$k=1$ theory with $N=3$ charged scalars. There are now three
vacua, given by $\sigma=m_i$ and $|q_j|^2=v^2\delta_{ij}$. In each
a $U(1)_{F_1}\times U(1)_{F_2}$ flavor symmetry is unbroken, under
which the scalars have charge
\be U(1)_{F_1}\times U(1)_{F_2}: \left\{\begin{array}{cl} q_1:& (+1,0) \\ q_2: &
(-1,1) \\ q_3: & (0,-1) \end{array}\right.\ee
The first elementary domain wall $\vec{g}=\vec{\alpha}_1$ interpolates between vacuum 1
and vacuum 2, breaking $U(1)_{F_1}$ along the way. The second elementary domain wall
interpolates between vacuum 2 and vacuum 3, breaking $U(1)_{F_2}$ along the way. Of interest
here is the domain wall $\vec{g}=\vec{\alpha}_1+\vec{\alpha}_2$ interpolating between
vacuum 1 and vacuum 3. It can be thought of as a composite of two domain walls.
The moduli space for these two domain walls
was studied in \cite{me,mmotw} and is of the form,
\be
{\cal W}_{\vg=\valpha_1+\valpha_2}\cong {\bf R}\times \frac{\R \times
\tilde{\cal W}_{\valpha_1+\valpha_2}}{\Z}
\label{12}\ee
%
\newcommand{\onefigurenocap}[1]{\begin{figure}[h]
         \begin{center}\leavevmode\epsfbox{#1.eps}\end{center}
         \end{figure}}
\newcommand{\onefigure}[2]{\begin{figure}[htbp]
         \begin{center}\leavevmode\epsfbox{#1.eps}\end{center}
         \caption{\small #2\label{#1}}
         \end{figure}}
\begin{figure}[tbp]
\begin{center}
\epsfxsize=3.0in\leavevmode\epsfbox{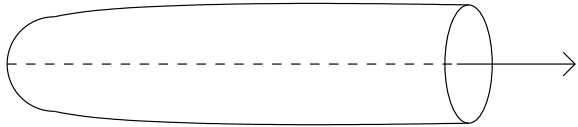}
\end{center}
\begin{center}
\begin{minipage}{13cm}
 \caption{\small 
The relative moduli space $\tilde{W}_{\valpha_1+\valpha_2}$ of two domain walls
is a cigar.}
\end{minipage}
\end{center}
\label{fig:reconnect}
\end{figure}
The first factor of $\R$ corresponds to the center of mass of the
two domain walls; the second factor corresponds to the combined
phase associated to the two domain walls. When the ratio of
tensions of the two elementary domain walls
$T_{\valpha_1}/T_{\valpha_2}$ is rational, the ratio of the
periods of the two phases are similarly rational and the second
$\R$ factor collapses to ${\bf S}^1$, while the quotient $\Z$
reduces to a finite group. The relative moduli space $\tilde{\cal
W}_{\valpha_1+\valpha_2}$ corresponds to the separation and
relative phases of the two domain walls. Importantly, and unlike
other solitons of higher co-dimension, the domain walls must obey
a strict ordering on the $x^3$ line: the $\vg=\valpha_1$ domain
wall must always be to the left of $\vg=\valpha_2$ domain wall.
The separation between the walls is therefore the halfline $\R^+$.
It was shown in \cite{me} that the relative phase is fibered over
$\R^+$ to give rise to a smooth cylinder, with the tip
corresponding to coincident domain walls. The resulting moduli
space is shown in Figure 1.

Note that the moduli space \eqn{12} is toric, inheriting two
isometries from the $U(1)_{F_1}\times U(1)_{F_2}$ symmetry. In an
abelian gauge theory with arbitrary number of flavors $N$, the
domain wall charge is always of the form $\vg=\sum_in_i\valpha_i$
with $n_i=0,1$, and the moduli space is always toric, meaning that
half of the dimensions correspond to $U(1)$ isometries.

\subsubsection*{\it Example 3: $\vg=\valpha_1+2\valpha_2+\valpha_3$}

In non-abelian theories, the domain wall moduli spaces are no
longer toric. The simplest such theory has a $U(2)$ gauge group
with $N=4$ fundamental scalars. The 6 vacua, and 15 different
domain walls, of this theory were detailed in \cite{boojum}. Under
the $U(1)^3_F$ flavor symmetry, the fundamental scalars transform
as
\be
U(1)_{F_1}\times U(1)_{F_2}\times U(1)_{F_3}:\left\{\begin{array}{cl}
q_1:& (+1,0,0) \\ q_2: & (-1,1,0) \\ q_3: & (0,-1,1) \\ q_4: & (0,0,-1)\\
\end{array}\right.
\label{3u1s}\ee
With this convention, the elementary domain wall $\vg=\valpha_i$ picks up its
phase from the action of the $U(1)_{F_i}$ flavor symmetry.

Here we concentrate on the domain wall system with the maximal
number of zero modes which arises from the choice of vacua
${\Xi}_-=(1,2)$ and $\Xi_+=(3,4)$ so that $\vg =
\valpha_1+2\valpha_2+\valpha_3$. This system can be separated into
four constituent domain walls. As explained in \cite{isonon,boojum},
the ordering of domain walls is no longer strictly
fixed in this example. The two outer elementary domain walls, on
the far left and far right, are each of the type $\vg=\valpha_2$.
However, the relative positions of the middle two domain walls,
$\vg=\valpha_1$ and $\valpha_3$ are not ordered and they may pass
through each other.

Unlike the situation for abelian gauge theories, the 8 dimensional
domain wall moduli space for this example is no longer toric;
${\cal W}_{\vg}$ inherits only three $U(1)$ isometries from
\eqn{3u1s}. Physically this means that the two phases associated
to the $\valpha_2$ domain walls are not both Goldstone modes and
they may interact as the domain walls approach. This behaviour
is familiar from the study of the Atiyah-Hitchin metric describing the
dynamics of two monopoles in $SU(2)$ gauge theory; we shall make
the analogy more precise in the following.

\subsection{The Ordering of Domain Walls}

As we stressed in the above examples, in contrast to other
solitons domain walls must obey some ordering on the line. This
will be an important ingredient when we come to extract domain
wall data from the linearized Nahm's equations in Section 4. Here
we linger to review this ordering.


The ordering of the elementary domain walls in non-abelian
theories was studied in detail in \cite{isonon}. One can derive
the result by considering the possible sequences of vacua as we
move over each domain wall. For example, we could consider the
``maximal domain wall'', interpolating between
$\Xi_-=\{1,2,\ldots,k\}$ and $\Xi_+=\{N-k+1,\ldots,N\}$. From the
left, the first elementary domain wall that we come across must be
$\vec{g}=\valpha_{k}$, corresponding to
$\Xi_-=(1,2,\ldots,k-1,k)\rightarrow (1,2,\ldots,k-1,k+1)$. The
next elementary domain wall may be either $\valpha_{k-1}$ or
$\valpha_{k+1}$. These two walls are free to pass through each
other, but cannot move further to the left than the $\valpha_k$
wall. And so on. Iterating this procedure, one finds that two
neighbouring elementary domain walls $\valpha_i$ and $\valpha_j$
may pass through each other whenever $\valpha_i\cdot\valpha_j=0$,
but otherwise have a fixed ordering on the line. The net result of
this analysis is summarized in Figure 2. The $x^3$ positions of
the domain walls are shown on the vertical axis; the position on
the horizontal axis denotes the type of elementary domain wall,
starting on the left with $\valpha_1$ and ending on the right with
$\valpha_{N-1}$.
\begin{figure}[tbp]
\begin{center}
\epsfxsize=14cm\leavevmode\epsfbox{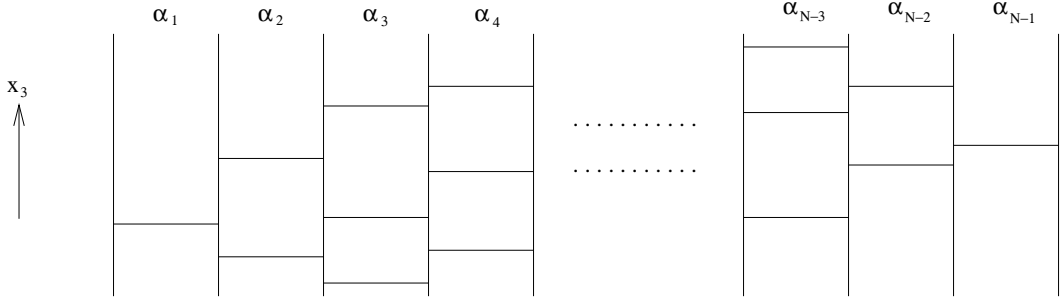}
\end{center}
\begin{center}
\begin{minipage}{15cm}
 \caption{\small
The ordering for the maximal domain wall. The $x^3$ spatial
direction is shown horizontally. The position in the vertical
direction denotes the type of domain wall. Domain walls of
neighbouring types have their positions interlaced.}
\end{minipage}
\end{center}
\label{fig:ninety   }
\end{figure}

In summary, we see that for $n_i=n_{i+1}-1$, the $\valpha_i$
domain walls are trapped between the $\valpha_{i+1}$ domain walls.
The reverse holds when $n_i=n_{i+1}+1$. Finally, when
$n_{i}=n_{i+1}$ the positions of the $\valpha_{i}$ domain walls
are interlaced with those of the $\valpha_{i+1}$ domain walls. The
last $\valpha_{i+1}$ domain wall is unconstrained by the
$\valpha_i$ walls in its travel in the positive $x^3$ direction,
although it may be trapped in turn by a $\valpha_{i+2}$ wall.

While we have discussed the maximal domain wall above, other sectors can
be reached either by removing some of the outer domain walls to infinity,
or by taking non-interacting products such subsets. It's important to note
that labelling a topological sector $\vg$ does not necessarily determine
the ordering of domain walls\footnote{An example: the
$\vg=\valpha_1+\valpha_2+\valpha_3$ domain wall. In the abelian theory with
$k=1$ and $N=4$, the ordering is $\valpha_1<\valpha_2<\valpha_3$. However,
in the non-abelian theory with $k=2$ and $N=4$, the ordering is
$\valpha_1,\valpha_3<\valpha_2$.}. We shall show that domain walls with the
same $\vg$, but different orderings, descend from different submanifolds
of the same monopole moduli space.

\section{Monopoles}

The main goal of this paper is to show how the moduli space of
domain walls introduced in the previous section is isomorphic to a
 submanifold of a related monopole moduli space. In this section
we review several relevant aspects of these monopole moduli
spaces.

It will turn out that
the domain walls of Section 2 are related to monopoles in an
$SU(N)$ gauge theory. Note that the flavor group from Section 2
has been promoted to a gauge group; we shall see the reason behind
this in Section 5. The Bogomoln'yi monopole equations are
\be B_\mu={\cal D}_\mu\phi\label{monbog}\ee
where $B_\mu$, $\mu=1,2,3$ is the $SU(N)$ magnetic field and $\phi$ is an adjoint
valued real scalar field. The monopoles exist only if $\phi$ takes a
vacuum expectation value,
\be \vev{\phi} ={\rm diag} (m_1,\ldots, m_{N})
\label{mono}\ee
where we take $m_i\neq m_j$ for $i\neq j$, ensuring breaking to
the maximal torus, $SU(N)\rightarrow U(1)^{N-1}$. It is not
coincidence that we've denoted the vacuum expectation values by
$m_i$, the same notation used for the masses in Section 2; it is
for this choice of vacuum that the correspondence holds.
(Specifically, the masses of the kinks will coincide with the masses
of monopoles, ensuring that the asymptotic metrics on ${\cal W}_{\vg}$
and ${\cal M}_{\vg}$ also coincide).

As described long ago by Goddard, Nuyts and Olive \cite{gno}, the
allowed magnetic charges under each unbroken $U(1)^{N-1}$ are
specified by a root vector\footnote{We ignore the factor of 2
difference between roots and co-roots. For simply laced groups,
such as $SU(N)$, it can be absorbed into convention.} of $su(N)$,
$\vg=(p_1,\ldots,p_{N})$. It is customary to decompose this in
terms of simple roots $\valpha_i$,
\be
\vec{g} =\sum_{i=1}^{N-1}n_i\vec{\alpha}_i\ \ \ \  \ \ , \ \ \ \ \ \ n_i\in\Z
\label{mg}\ee
Once again, the notation is identical to that used for domain
walls \eqn{dwg} for good reason. Solutions to the monopole
equations \eqn{monbog} exist for all values of $n_i\geq 0$.
This is in contrast to domain walls where, as we have seen, configurations
only exist in a finite number of sectors defined by $p_i=0$ or
$p_i=\pm 1$.  The mass of the magnetic monopole is
$M_{\rm mono}=(2\pi/e^2)\vec{m}\cdot\vg$.

The monopole moduli space ${\cal M}_{\vg}$ is the space of solutions to \eqn{mono}
in a fixed topological sector $\vg$. The dimension of this space, equal to the number
of zero modes of given solution, was computed by E. Weinberg in \cite{erick} using
Callias' version of the index theorem. The result is:
\be {\rm dim}\,({\cal M}_{\vg})=4\sum_{i=1}^{N-1}n_i
\label{mondim}\ee
which is to be compared with \eqn{dwdim}.

\subsection{The Relationship between ${\cal M}_{\vg}$ and ${\cal W}_{\vg}$}

We are now in a position to describe the relationship between the moduli space of
domain walls ${\cal W}_{\vg}$ and the moduli space of magnetic monopoles ${\cal M}_{\vg}$.
We will show that ${\cal W}_{\vg}$ is a complex, middle dimensional, submanifold of
${\cal M}_{\vg}$, defined by the fixed point set of the action rotating the monopoles
in a plane, together with a suitable gauge action.
To do this, we first need to describe the symmetries of ${\cal M}_{\vg}$.

The monopole moduli space ${\cal M}_{\vg}$ admits a natural,
smooth, hyperK\"ahler metric \cite{manton,ah}. For generic $\vg$
this metric enjoys $(N-1)$ tri-holomorphic isometries arising from
the action of the $U(1)^{N-1}$ abelian gauge group. Further the
metric has an $SU(2)_R$ symmetry, arising from rotations of the
monopoles in $\R^3$, which acts on the three complex structures of
${\cal M}_{\vg}$. In other words, any $U(1)_R\subset SU(2)_R$ is a
holomorphic isometry, preserving a single complex structure while
revolving the remaining two. Let us choose $U(1)_R$ to rotate the
monopoles in the $(x^2-x^3)$ plane. In what follows we will be
interested in a specific holomorphic $\hat{U}(1)$ action which
acts simultaneously by a $U(1)_R$ rotation and a linear
combination of the gauge rotations $U(1)^{N-1}$ (to be specified
presently). We denote the Killing vector on ${\cal M}_{\vg}$
associated to $\hat{U}(1)$ as $\hat{k}$. We claim
\be {\cal W}_{\vg}=\left.{\cal M}_{\vg}\right|_{\hat{k}=0}
\label{result}\ee
This result holds at the level of topology and asymptotic metric
of the spaces. The manifold ${\cal W}_{\vg}$ inherits a metric
from ${\cal M}_{\vg}$ by this reduction: it does not coincide with
the domain wall metric in the interior on ${\cal W}_{\vg}$. (For
example, corrections to the asymptotic metric on ${\cal W}_{\vg}$ are
exponentially suppressed while those of ${\cal M}_{\vg}$ have
power law behaviour). It would be interesting to examine if ${\cal
W}_{\vg}$ inherits the correct K\"ahler class and/or complex
structure from ${\cal M}_{\vg}$.

We defer a derivation of \eqn{result} to the following section, but first present
some simple examples.

\subsubsection*{\it Example 1: $\vg=\valpha_1$}

Monopoles in $SU(2)$ gauge theories are labelled by a single topological
charge $\vg=n_1\valpha_1$. For a single monopole ($n_1=1$) the moduli
space is simply
\be
{\cal M}_{\vg=\valpha_1}\cong\R^3\times {\bf S}^1
\ee
where the $\R^3$ factor denotes the position of the monopole,
while the ${\bf S}^1$ arises from global gauge transformations
under the surviving $U(1)$. The radius of the ${\bf S}^1$ is
$2\pi /(m_1-m_2)$. In this case the $\hat{U}(1)$ action coincides with
the rotation $U(1)_R$ in the $(x^2-x^3)$ plane and we have
trivially
\be
{\cal W}_{\valpha_1}\cong \R\times {\bf S}^1\cong
\left.{\cal M}_{\valpha_1}\right|_{\hat{k}=0}
\ee
The similarity between the domain wall and monopole moduli spaces
for a single soliton was noted by Abraham and Townsend
\cite{edpaul}. In both cases, motion in the ${\bf S}^1$ factor
gives rise to dyonic solitons.

Note that monopole moduli spaces for charges $\vg=n_1\valpha_1$
exist for all $n_1\in\Z^+$. For example, the $n_1=2$ monopole
moduli space is home to the famous Atiyah-Hitchin metric
\cite{ah}. However, there is no domain wall moduli space with this
charge in the class of theories we discuss in Section 2.

\subsubsection*{\it Example 2: $\vg=\valpha_1+\valpha_2$}

Our second example is the $\vg=\valpha_1+\valpha_2$ monopole in
$SU(3)$ gauge theories (sometimes referred to as the $(1,1)$ monopole). The
moduli space was determined in \cite{conell,gl,lwy1} to be of the
form
\be
{\cal M}_{\vg=\valpha_1+\valpha_2}\cong\R^3\times\frac{\R\times
\tilde{\cal M}_{\valpha_1+\valpha_2}}{\Z}
\ee
where the relative moduli space $\tilde{\cal M}_{\valpha_1+\valpha_2}$ is the
Euclidean Taub-NUT space, endowed with the metric
\be
ds^2=\left(1+\frac{1}{r}\right)\,(dr^2+r^2d\theta^2+^2\sin^2\theta\,d\phi^2)
+\left(1+\frac{1}{r}\right)^{-1}(d\psi+\cos\theta d\phi)^2
\label{tn}\ee
Here $r$,$\theta$ and $\phi$ are the
familiar spherical polar coordinates. The coordinate $\psi$ arises
from $U(1)$ gauge transformations. The manifold has a $SU(2)_R\times U(1)$
isometry, of which only a $U(1)_R\times U(1)$ are manifest in the above
coordinates. The holomorphic $U(1)_R$ isometry acts by rotating the two
monopoles: $\phi\rightarrow \phi+c$. The tri-holomorpic $U(1)$ isometry
changes the relative phase of the monopoles: $\psi\rightarrow \psi+c$.
Both of these actions have a unique fixed point at $r=0$, the ``nut'' of Taub-NUT.
However, the combined action with Killing vector $\partial_\psi+\partial_\phi$
has a fixed point along the half-line $\theta=\pi$, with $\psi$ fibered over this
line to produce the cigar shown in Figure 1. This is the relative moduli space
$\tilde{\cal W}_{\valpha_1+\valpha_2}$.

Similar calculations hold for monopoles of charge
$\vg=\sum_{i=1}^{N-1}\valpha_i$, whose dynamics is described by a
class of toric hyperK\"ahler metrics, known as the Lee-Weinberg-Yi
metrics \cite{lwy2}. Once again, a suitable ${\bf S}^1$ action on
these spaces can be identified such that the fixed points localize
on ${\cal W}_{\vg}$, the moduli space of domain walls in $U(1)$
gauge theories with $N$ charged scalars.

\subsubsection*{\it Example 3: $\vg=\valpha_1+2\valpha_2+\valpha_1$}

As described in the previous section, the simplest domain wall charge
$\vg=\sum_in_i\valpha_i$ with some $n_i>1$ occurs for $\vg=\valpha_1+
2\valpha_2+\valpha_3$, and corresponds to a monopole in a $SU(4)$
gauge theory. No explicit expression for the  metric on this monopole moduli space is
known although, given the results of \cite{21}, such a computation may be
feasible. Without an explicit expression for the metric in this, and more
complicated examples, we need a more powerful method to describe the
moduli space. This is provided by the Nahm construction,
which we now review

\subsection{D-Branes and Nahm's Equations}

The moduli space of magnetic monopoles is isomorphic to the moduli
space of Nahm data. Here we review the Nahm construction of the
monopole moduli space \cite{nahm} and, in particular, the
embedding within the framework of D-branes due to Diaconescu
\cite{diac}. This will be useful to compare to the domain walls of
the next section.

In the D-brane construction, Nahm's equations arise as the
low-energy description of D-strings suspended between D3-branes
\cite{diac}. The $SU(N)$ Yang-Mills theory lives on the
worldvolume of $N$ D3-branes separated in, say, the $x_6$
direction, with the $i^{\rm th}$ D3-brane placed at position
$(x_6)_i=m_i$ in accord with the adjoint expectation value
\eqn{mono}. The monopole of charge $\vg=\sum_in_i\valpha_i$
corresponds to suspending $n_i$ D-strings between the $i^{\rm th}$
and $(i+1)^{\rm th}$ D3-brane. This configuration is shown in
Figure 3.
\begin{figure}[tbp]
\begin{center}
\epsfxsize=10cm\leavevmode\epsfbox{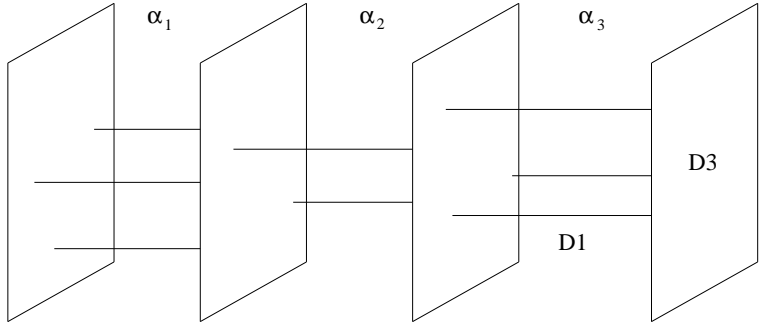}
\end{center}
\begin{center}
\begin{minipage}{15cm}
 \caption{\small
The $\vg=3\valpha_1+2\valpha_2+3\valpha_3$ monopole as D-strings
stretched between D3-branes.}
\end{minipage}
\end{center}
\end{figure}

The motion of the D-strings in each segment $m_i\leq x_6\leq m_{i+1}$ is
governed by four hermitian $n_i\times n_i$  matrices,
$X_1,X_2,X_3$ and $A_6$ subject to the covariant version of Nahm's equations,
\be
\frac{dX_\mu}{dx_6}-i[A_6,X_\mu]-\frac{i}{2}\epsilon_{\mu\nu\rho}
[X_\nu,X_\rho]=0\ \ \ \ \ \ \ \ m_i\leq x_6\leq m_{i+1}
\label{nahm}\ee
%
modulo $U(n_i)$ gauge transformations acting on the interval $m_i\leq
x_6\leq m_{i+1}$, and vanishing  at the boundaries. The $X_\mu$ form the
triplet representation of the $SU(2)_R$ symmetry which rotates monopoles in $\R^3$.
The $U(1)^{N-1}$ surviving gauge transformations acts on the Nahm data by constant
shifts of the $(N-1)$ ``Wilson lines'' $A_6\rightarrow A_6 + c1_{n_i}$.

The interactions
between neighbouring segments depends on the relative size of
the matrices and is given by \cite{hub}

\underline{\ $n_i=n_{i+1}$:\ } In this case the $U(n_i)$ gauge symmetry is
extended to the interval $m_i\leq x_6\leq m_{i+2}$ and an impurity is
added to the right-hand-side of Nahm's equations, which now read
\be
\frac{dX_\mu}{dx_6}-i[A_6,X_\mu]-\frac{i}{2}\epsilon_{\mu\nu\rho}
[X_\nu,X_\rho]=\omega_\alpha\, \sigma_\mu^{\alpha\beta}\,\omega^\dagger_\beta\
\delta(x_6-m_{i+1})
\label{imp}\ee
Here $\sigma_\mu$ are the Pauli matrices. The impurity degrees of freedom
lie in the complex 2-vector, $\omega_\alpha=(\psi,\tilde{\psi}^\dagger)$ which
is a doublet under the $SU(2)_R$ symmetry. Both $\psi$ and $\tilde{\psi}^\dagger$
are themselves complex $n_i$
vectors, transforming in the fundamental representation of the $U(n_i)$
gauge group. The combination
$\omega_\alpha \sigma_\mu^{\alpha\beta}\omega^\dagger_\beta$ is thus
an $n_i\times n_i$
matrix, transforming in the adjoint representation of the gauge group.
The $\omega_\alpha$ fields can be thought of as a hypermultiplet
arising from $D1-D3$ strings \cite{hw,kapset,tsimpis}

\underline{\ $n_i=n_{i+1}-1$:\ } In this case $X_\mu\rightarrow (X_\mu)_-$, a set
of three $n_i\times n_i$ matrices, as $x_6\rightarrow (m_i)_-$ from the left. To
the right of $m_i$, the $X_\mu$ are $(n_i+1)\times (n_i+1)$ matrices which
must obey
\be
X_\mu\rightarrow \left(\begin{array}{cc} y_\mu & a_\mu^\dagger \\ a_\mu & (X_\mu)_-
\end{array}\right)\ \ \ \ \ {\rm as\ } x_6\rightarrow (m_i)_+
\ee
where $y_\mu\in\R$ and each $a_\mu$ is a complex $n_i$-vector.

\underline{\ $n_i\leq n_{i+1}-2$:\ } Once again we take $X_\mu\rightarrow
(X_\mu)_-$ as $x_6\rightarrow (m_i)_-$ but, from the other side, the
matrices $X_\mu$ now have a simple pole at the boundary,
\be
X_\mu\rightarrow \left(\begin{array}{cc} J_\mu/(x_6-m_i)+ Y_\mu &
0 \\ 0 & (X_\mu)_- \end{array}\right)\ \ \ \ \ {\rm as\ }
x_6\rightarrow (m_i)_+
\ee
where $J_\mu$ is the irreducible $(n_{i+1}-n_i)\times (n_{i+1}-n_i)$
representation of $su(2)$, and $Y_\mu$ are now constant
$(n_{i+1}-n_i)\times (n_{i+1}-n_i)$ matrices.

Case 2 above is usually described as a subset of Case 3 (with the
one-dimensional irreducible $su(2)$ representation given by $J_\mu=0$).
Here we have listed Case 2 separately since when we come to describe
a similar construction for domain walls, only Case 1 and 2 above will
appear. The conditions for $n_i<n_{i+1}$ were derived in \cite{chenwein}
by starting with the impurity data \eqn{imp} and taking several monopoles
to infinity. Obviously, for $n_i>n_{i+1}$, one imposes the same boundary
conditions described above, only flipped in the $x_6$ direction.

The space of solutions to Nahm's equations, subject to the
boundary conditions detailed above, is isomorphic to the monopole
moduli space ${\cal M}_{\vg}$. Moreover, there exists a natural
hyperK\"ahler metric on the solutions to Nahm's equations which
can be shown to coincide with the Manton metric on the monopole
moduli space. For the $\vg=\valpha_1+\valpha_2$ monopole, the
metric on the associated Nahm data was computed in \cite{conell}
and shown to give rise to the Euclidean Taub-NUT metric \eqn{tn}.
For the $\vg=\sum_i\valpha_i$ monopoles, the corresponding
computation was performed in \cite{murray}, resulting in the
Lee-Weinberg-Yi metrics \cite{lwy2}.

\section{D-Branes and Domain Walls}

In this section we would like to realize the domain walls that we
described in Section 2 on the worldvolume of D-branes, mimicking
Diaconescu's construction for monopoles.  From the resulting
D-brane set-up we shall read off the world-volume dynamics of the
domain walls to find that they are described by a truncated
version of Nahm's equations \eqn{nahm}. Nahm's equations have also
arisen as a description of domain walls in ${\cal N}=1^\star$
theories \cite{boris}, although the relationship, if any, with the
current work is unclear. Domain walls of the type described in
Section 2 were previously embedded in D-branes in
\cite{kinky,ohta} and several properties of the solitons were
extracted (see in particular the latter reference). However, the
worldvolume dynamics of the walls is difficult to determine in
these set-ups and the relationship to magnetic monopoles obscured.

\begin{figure}[tbp]
\begin{center}
\epsfxsize=15cm\leavevmode\epsfbox{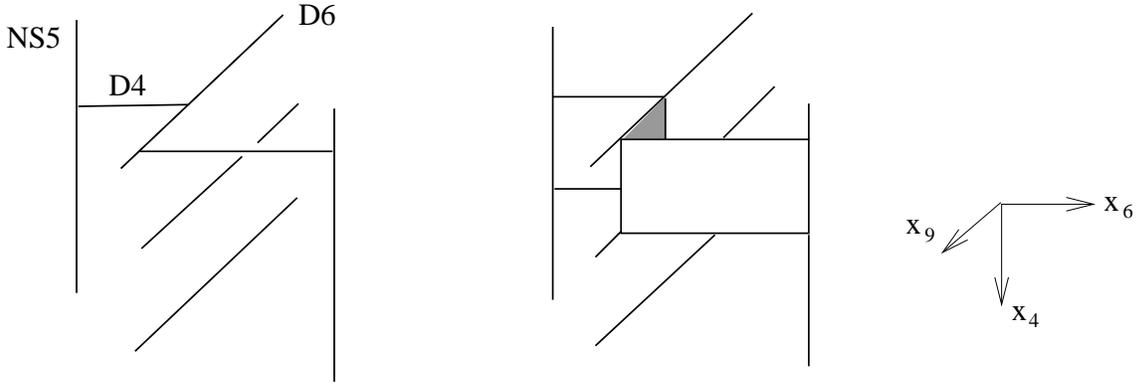}
\end{center}
\begin{center}
\begin{minipage}{15cm}
 \caption{\small
The D-brane set-up for the $U(1)$ gauge theory with $N=3$ flavors. The
vacuum is shown on the left; the elementary domain wall $\vg=\valpha_1$
on the right.}
\end{minipage}
\end{center}
\end{figure}

We start by constructing the theory with eight supercharges on the
worldvolume of D-branes \cite{hw}. For definiteness we choose to
build the ${\cal N}=2$, $d=3+1$ theory in IIA string theory
although, by T-duality, we could equivalently work with any
spacetime dimension\footnote{In fact, as explained in
\cite{witten}, the overall $U(1)\subset U(k)$ is decoupled in
the IIA brane set-up after lifting to M-theory. This effect will
not concern us here.}. The construction is well known and is drawn
in Figure 4. We suspend $k$ D4-branes between two NS5-branes,
and insert a further $N$ D6-branes to play the role of the
fundamental hypermultiplets. The worldvolume dimensions of the
branes are
\be NS5: && 012345 \nn\\ D4: && 01236 \nn\\ D6: && 0123789 \nn\ee
The gauge coupling $e^2$ and the Higgs vev $v^2$ are encoded in
the separation of the NS5-branes in the $x_6$ and $x_9$ directions
respectively, while the masses $m_i$ are determined by the
positions of the D6-branes in the $x_4$ direction (we choose the
D6-branes to be coincident in the $x_5$ direction, corresponding
to choosing all masses to be real).
\be \frac{1}{e^2} \sim \left.\frac{\Delta
x_6}{l_sg_s}\right|_{NS5}\ \ \  \ ,\ \ \ \ v^2 \sim
\left.\frac{\Delta x_9}{l_s^3g_s}\right|_{NS5}\ \ \ \ , \ \ \ \
m_i\sim -\left.\frac{x_4}{l_s^2}\right|_{D6_i}\ee
After turning on the Higgs vev $v^2$, the D4-branes must split on
the D6-branes in order to preserve supersymmetry. The S-rule
\cite{hw} ensures that each D6-brane may play host to only a
single D4-brane. In this manner a vacuum of the theory is chosen
by picking $k$ out of the $N$ D6-branes on which the D4-branes
end, in agreement with equation \eqn{set}.

The domain walls correspond to a configuration of D4-branes which
start life at $x^3=-\infty$ in a vacuum configuration ${\Xi}_-$,
and end up at $x^3=+\infty$ in a distinct vacuum $\Xi_+$. As is
clear from Figure 4, as D4-branes walls interpolate in $x^1$, they
must also move in both the $x^4$ direction and the $x^9$ direction
\cite{hh}. The NS-branes and D6-branes are linked, meaning that a
D4-brane is either created or destroyed as they pass the
NS5-branes in the $x^6$ direction \cite{hw}. In the domain wall
background, which of these possibilities occurs differs if we move
the D6-branes to the left or right since the D4-brane charge
varies from one end of the domain wall to the other.

As it stands, it is difficult to read off the dynamics of the
D4-branes in Figure 4. However, we can make progress by taking the
$e^2\rightarrow \infty$ limit, in which the two NS5-branes become
coincident in the $x^6$ direction. After rotating our viewpoint,
the system of branes now looks like the ladder configuration shown
in Figure 5 (note that we have also rotated the branes relative to
Figure 4, so the horizontal is the $x^4$ direction). We are left
with a series of D4-branes, now with worldvolume $02349$,
stretched between $N$ D6-branes, while simultaneously sandwiched
between two NS5-branes. Following these manoeuvres, one finds that
the domain wall $\vg=\sum_in_i\valpha_i$ results in $n_i$
D4-branes stretched between the $i^{\rm th}$ and $(i+1)^{\rm th}$
D6-branes (counting from the top, since we
have chosen the ordering $m_i > m_{i+1}$.

\begin{figure}[tbp]
\begin{center}
\epsfxsize=15cm\leavevmode\epsfbox{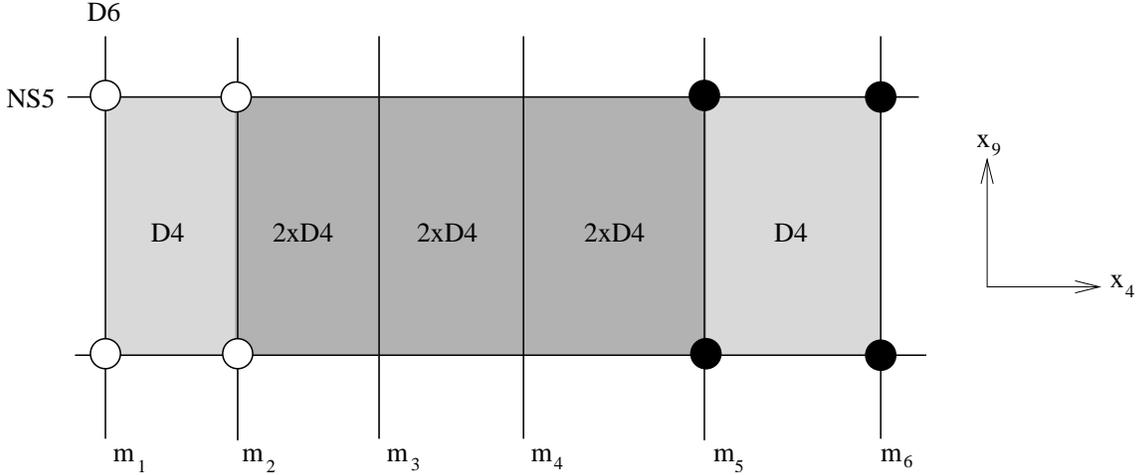}
\end{center}
\begin{center}
\begin{minipage}{15cm}
 \caption{\small
The D-brane set-up for the $U(2)$ gauge theory with $N=6$ flavors. The
maximal
$\vg=\valpha_1+\valpha_5+2(\valpha_2+\valpha_3+\valpha_4)$ domain wall
is shown.}
\end{minipage}
\end{center}
\end{figure}

It may be worth describing how the domain wall charges arise directly
in the set-up of Figure
5. We start in a chosen vacuum $\Xi_-$, denoted
by placing $k$ pairs of white dots on $N$ distinct D6-branes, as
shown in the figure. A domain wall arises every time a pair of dots
is proceeds to another D6-brane, dragging a D4-brane behind it like
clingwrap. The S-rule translates to the fact that two pairs of dots may not
simultaneously lie on the same D6-brane. The final vacuum $\Xi_+$ is
denoted by the black dots in the figure and the domain wall charges
$n_i$ are given by the number of times a D4-brane has been pulled between
the $i^{\rm th}$ and $(i+1)^{\rm th}$ D6-branes.

\subsection{Domain Wall Dynamics}

We are now in a position to read off the dynamics of the domain
walls. In the absence of the NS5-branes, the D4-branes would
stretch to infinity in the $x_9$ direction, and the resulting
D-brane set-up in Figure 5 is T-dual to the monopoles in Figure 3.
The presence of the NS5-branes projects out half the degrees of
freedom of the monopoles, leaving a simple linear set of
equations. In each segment $m_i\leq x_4\leq m_{i+1}$ the domain
walls are described by two $n_i\times n_i$ matrices $X_3$ and
$A_4$ satisfying
\be \frac{dX_3}{dx_4}-i[A_4,X_3]=0 \label{dwnahm}\ee
modulo $U(n_i)$ gauge transformations acting on the interval $m_i\leq
x_4\leq m_{i+1}$, and vanishing  at the boundaries. As in the case of
monopoles, the interactions between neighbouring segments depends on the
relative size of the matrices:

\underline{\ $n_i=n_{i+1}$:\ } Again, the $U(n_i)$ gauge symmetry is
extended to the interval $m_i\leq x_4\leq m_{i+2}$ and an impurity is
added to the right-hand-side of Nahm's equations, which now read
\be \frac{dX_3}{dx_4}-i[A_4,X_3]=
\pm\psi\psi^\dagger\delta(x_4-m_{i+1}) \label{dwimp}\ee
where the impurity degree of freedom $\psi$ transforms in the
fundamental representation of the $U(n_i)$ gauge group, ensuring
the combination $\psi\psi^\dagger$ is a $n_i\times n_i$ matrix
transforming, like $X_3$, in the adjoint representation. These
$\psi$ degrees of freedom are chiral multiplets which survive the
NS5-brane projection. We shall see shortly that the choice of
$\pm$ sign will dictate the relative ordering of the domain walls
along the $x_3$ direction.

\underline{\ $n_i=n_{i+1}-1$:\ } In this case $X_3\rightarrow
(X_3)_-$, an $n_i\times n_i$ matrix, as $x_4\rightarrow (m_i)_-$
from the left. To the right of $m_i$, $X_3$ is a $(n_i+1)\times
(n_i+1)$ matrix obeying
\be X_3\rightarrow \left(\begin{array}{cc} y & a^\dagger \\ a &
(X)_-
\end{array}\right)\ \ \ \ \ {\rm as\ } x_4\rightarrow (m_i)_+
\label{bc}\ee
where $y_\mu\in\R$ and each $a_\mu$ is a complex $n_i$-vector. The obvious analog
of this boundary condition holds when $n_i=n_{i+1}+1$.

These boundary conditions obviously descend from the original Nahm
boundary conditions for monopoles. Just as the space of Nahm data
is isomorphic to the moduli space of magnetic monopoles, we
conjecture that the moduli space of linearized Nahm data described
above is isomorphic to the moduli space of domain walls. We shall
shortly show that it indeed captures the most relevant aspect of
domain walls: their ordering. In fact, the linearized Nahm
equations \eqn{dwnahm} are rather trivial to solve. We first
employ the $\prod_iU(n_i)$ gauge transformations to make
$A_4(x_4)$ a constant in each interval $m_i\leq x_4\leq m_{i+1}$.
This can be achieved by first diagonalizing $A_4$, and
subsequently acting with the $U(1)^{n_i}$ transformation
$A_4\rightarrow A_4-\partial_4\alpha$ where, in each segment
$m_i\leq x_4\leq m_{i+1}$, $\alpha$ is given by
\be \alpha(x_4)=\int_{m_i}^{x_4}\ A_4(x^\prime_4)\,dx_4^\prime -\left[\int_{m_i}^{m_{i+1}}
A_4(x^\prime_4)\,dx^\prime_4\right]\frac{m_i-x_4}{m_i-m_{i+1}}
\label{gt}\ee
which has the property that $\alpha(m_i)=\alpha(m_{i+1})=0$.
Further gauge transformations with non-zero winding on the
interval ensure that $A_4$ is periodic, with each eigenvalue lying
in $A_4\in [0,2\pi /(m_i-m_{i+1}))$. These $N-1$ ``Wilson lines''
will play the role of the phases associated to domain wall system.
Note that when $n_i=n_{i+1}$, the above choice of gauge leaves a
residual $U(n_i)$ gauge symmetry acting only on the chiral
impurity $\psi$. In this gauge we can now easily integrate
\eqn{dwnahm} in each interval,
\be X_3(x_4)=e^{iA_4x_4}\,\hat{X}_3\,e^{-iA_4x_4} \ee
where the eigenvalues of $X_3$ are independent of $x_4$ in each
interval. We identify these $n_i$ eigenvalues with the positions
of the $n_i$ $\valpha_i$ elementary domain walls.

We are now in a position to derive the linearized Nahm equations
\eqn{dwnahm} from the original Nahm equations \eqn{nahm} in terms
of a fixed point set of a $\hat{U}(1)$ action. Consider first the
action of the $U(1)_R\subset SU(2)_R$ isometry on the Nahm data,
which rotates $X_1$ and $X_2$ while leaving $X_3$ fixed. This
rotation also acts on the impurity
$\omega=(\psi,\tilde{\psi}^\dagger)$ by $(\psi,
\tilde{\psi})\rightarrow e^{i\alpha}(\psi,\tilde{\psi})$. To
retain half of the impurities for the domain wall equations
\eqn{dwimp}, we need to compensate for this transformation with
the residual $U(1)\subset U(n_i)$ transformation acting on the
appropriate impurity $\omega$ by $\omega\rightarrow
e^{i\beta}\omega$. By choosing $\beta=\pm\alpha$ we can pick a
$\hat{U}(1)$ action which leaves either the $\psi$ or the
$\tilde{\psi}$ impurity invariant. Which we choose to save is
correlated with the choice of minus sign in \eqn{dwimp} which, in
turn, dictates the ordering of neighbouring domain walls as we
shall now demonstrate.

To summarize, we have shown the that description of domain wall
dynamics \eqn{dwnahm} arises from the fixed point of a
$\hat{U}(1)$ on the original Nahm equations \eqn{nahm}. This
action descends to a $\hat{U}(1)$ isometry on the monopole moduli
space ${\cal M}_{\vg}$, the fixed points of which coincide with
the domain wall moduli space ${\cal W}_{\vg}$. A physical explanation
for this correspondence follows along the lines of \cite{stillme}:
in theories in the Higgs phase, confined magnetic monopoles with charge
$\vg$ exist, emitting $k$ multiple vortex strings. When
these vortex strings coincide, the worldvolume theory is of the form
described in Section 2 \cite{vib} and the monopoles appear as charge
$\vg$ kinks.

\subsection{The Ordering of Domain Walls Revisited}

As explained in Section 2, in contrast to monopoles, domain walls
must satisfy a specific ordering on the $x^3$ line. We will now
show that this ordering is encoded in the boundary conditions
described above. Suppose first that $n_i=n_{i+1}$. The positions
of the $\valpha_i$ domain walls are given by the eigenvalues of
$X_3$ restricted to the interval $m_i\leq x_4\leq m_{i+1}$. Let us
denote this matrix as $X_3^{(i)}$ and the eigenvalues as
$\lambda^{(i)}_m$, where $m=1,\ldots n_i$.  The impurity
\eqn{dwimp} relates the two sets of eigenvalues by the jumping
condition
\be X_3^{(i+1)}=X_3^{(i)}+\psi\psi^\dagger \label{jump}\ee
where we have chosen the positive sign for definiteness.  However,
from the discussion in Section 2 (see, in particular, figure 3) we
know that the domain walls cannot have arbitrary position but must
be interlaced,
\be
\lambda^{(i)}_1\leq \lambda^{(i+1)}_1\leq\lambda^{(i)}_2\leq\ldots\leq\lambda^{(i+1)}_{n_i-1}
\leq\lambda^{(i)}_{n_i}\leq\lambda^{(i+1)}_{n_i}
\label{order}\ee
We will now show that the ordering of domain walls \eqn{order} follows from
the impurity jumping condition \eqn{jump}.

To see this, consider firstly the situation in which
$\psi^\dagger\psi\ll \Delta\lambda^{(i)}_{m}$ so that the matrix
$\psi\psi^\dagger$ may be treated as a small perturbation of
$X_3^{(i)}$. The positivity of $\psi\psi^\dagger$ ensures that
each $\lambda_m^{(i+1)}\geq \lambda_m^{(i)}$. Moreover, it is
simple to show that the $\lambda_m^{(i+1)}$ increase monotonically
with $\psi^\dagger\psi$. This leaves us to consider the other
extreme, in which $\psi^\dagger\psi\rightarrow \infty$. It this
limit $\psi$ becomes one of the eigenvectors of $X_3^{(i+1)}$ with
eigenvalue $\lambda^{(i+1)}_{n_i}=\psi^\dagger\psi\rightarrow
\infty$ which reflects the fact that this limit corresponds to the
situation in which the last domain wall is taken to infinity. What
we want to show is that the remaining $(n_i-1)$ $\valpha_{i+1}$
domain walls are trapped between the $n_i$ $\valpha_i$ domain
walls as depicted in Figure 3. Define the $n_i\times n_i$
projection operator
\be
P=1-\hat{\psi}\hat{\psi}^\dagger
\ee
where $\hat{\psi}=\psi/\sqrt{\psi^\dagger\psi}$. The positions of
the remaining $(n_i-1)$ $\valpha_{i+1}$ domain walls are given by
the (non-zero) eigenvalues of $PX_3^{(i)}P$. We must show that,
given a rank $n$ hermitian matrix $X$, the eigenvalues of $PXP$
are trapped between the eigenvalues of $X$. This elementary
property of hermitian matrices can be seen as follows:
\be
\det(PXP-\mu)&=&\det(XP-\mu) \nn\\
&=&\det(X-\mu-X\hat{\psi}\hat{\psi}^\dagger)\nn\\ &=&\det(X-\mu)
\det(1-(X-\mu)^{-1}X\hat{\psi}\hat{\psi}^\dagger)\nn\ee
Since $\hat{\psi}\hat{\psi}^\dagger$ is rank one, we can write this as
\be
\det(PXP-\mu)&=&
\det(X-\mu)\,[1-\Tr((X-\mu)^{-1}X\hat{\psi}\hat{\psi}^\dagger)]\nn\\
&=& -\mu\,\det(X-\mu)\,\Tr((X-\mu)^{-1}\hat{\psi}\hat{\psi}^\dagger)
\nn\\ &=& -\mu\left[\prod_{m=1}^n(\lambda_m-\mu)\right]\,\left[\sum_{m=1}^n
\frac{|\hat{\psi}_m|^2}{\lambda_m-\mu}\right]
\ee
where $\hat{\psi}_m$ is the $m^{\rm th}$ component of the vector $\psi$.
We learn that $PXP$ has one zero eigenvalue while, if the eigenvalues
$\lambda_m$ of $X$ are distinct, then the eigenvalues of $PXP$ lie at the
roots the function
\be R(\mu)=\sum_{m=1}^n\frac{|\hat{\psi}_m|^2}{\lambda_m-\mu}\ee
The roots of $R(\mu)$ indeed lie between the eigenvalues $\lambda_m$. This
completes the proof that the impurities \eqn{dwimp} capture the correct ordering
of the domain walls.

The same argument shows that the boundary condition \eqn{bc} gives rise to the
correct ordering of domain walls when $n_{i+1}=n_i+1$, with the $\valpha_{i}$
domain walls interlaced between the $\valpha_{i+1}$ domains walls. Indeed, it
is not hard to show that \eqn{bc} arises from \eqn{dwimp} in the limit that
one of the domain walls is taken to infinity.

\section*{Acknowledgement}
AH is supported in part by the CTP and LNS of MIT, DOE contract
$\#$DE-FC02-94ER40818, NSF grant PHY-00-96515, the BSF
American-Israeli Bi-national Science Foundation and a DOE OJI
Award.  DT is supported by the Royal Society.

\end{document}